\documentclass[prd,nofootinbib]{revtex4}

\usepackage{graphicx}
\usepackage{amsmath}
\usepackage{amssymb}
\usepackage{hyperref}
\usepackage{bm}
\usepackage{ulem}
\usepackage{cleveref}
\usepackage{bbm}

\newcommand{\nc}{\newcommand}
\nc{\beq}{\begin{equation}}  \nc{\eeq}{\end{equation}}
\nc{\bea}{\begin{eqnarray}}  \nc{\eea}{\end{eqnarray}}
\nc{\baa}{\begin{array}}     \nc{\eaa}{\end{array}}
\nc{\bit}{\begin{itemize}}   \nc{\eit}{\end{itemize}}
\nc{\ben}{\begin{enumerate}} \nc{\een}{\end{enumerate}}
\nc{\bce}{\begin{center}}    \nc{\ece}{\end{center}}
\nc{\bpm}{\begin{pmatrix}}   \nc{\epm}{\end{pmatrix}}
\nc{\bvt}{\begin{verbatim}}  \nc{\evt}{\end{verbatim}}

\def\gev{\hbox{GeV}}
\def\ev{\hbox{eV}}
\def\ui{U(1)}
\def\lcal{{\cal L}}
\def\half{\frac12}
\def\vev{vacuum expectation value}
\def\vevof#1{\left\langle #1 \right\rangle}
\def\lesim{\lesssim}
\def\pp{{\bf p}}
\def\up#1{^{\left( #1 \right) }}
\def\inv#1{\frac1{#1}}
\def\lhs{left hand side\ }
\def\kev{\hbox{keV}}
\def\gesim{\gtrsim}
\def\then{{\quad\Rightarrow\quad}}
\def\tev{\hbox{TeV}}
\def\mev{\hbox{MeV}}
\def\ocal{{\cal O}}
\def\hBB{{\mathbbm H}}
\def\xx{{\bf x}}
\def\kk{{\bf k}}
\def\qq{{\bf q}}
\def\gcal{{\cal G}}
\def\ket#1{\left| #1 \right\rangle}

\def\tBB{{\mathbbm T}}
\def\vcal{{\cal V}}
\def\PP{{\bf P}}
\def\cm{\hbox{cm}}
\def\pibf{{\bm\pi}}	
\def\fcal{{\cal F}}
\def\deriva#1#2#3{\left(\frac{\partial #1}{\partial #2}\right)_{#3}}
\def\rBB{{\mathbbm R}}
\def\mati{{\mathbbm1}}
\def\ecal{{\cal E}}
\def\AA{{\bf A}}

\def\bal{\begin{align}}
\def\mcr{\nonumber\\[6pt]} 

\def\tr#1{{\rm tr}\left\{ #1 \right\}}

\def\fBB{{\mathbbm F}}

\def\dpp{\frac{d^3\pp}{(2\pi)^3}}

\def\be{_{\tt be}}
\def\sm{_{\tt sm}}
\def\tot{_{\tt tot}}

\def\ch{q\be}
\def\Ch{Q\be}
\def\mbe{{m\be}}
\def\lbe{\lambda\be}
\def\ssm{s\sm}
\def\sbe{s\be}
\def\rsm{\rho\sm}

\def\bec{C}

\def\gs{g_\star}
\def\gss{g_{\star{\tt s}}}

\def\x{{\tt x}}
\def\eg{u_\Gamma} 

\def\dt#1{d\tilde{\bf#1}\,}
\def\dtp {\dt p}
\def\dtq {\dt q}
\def\dtk {\dt k}

\def\bfz{{\bf{0}}}

\def\msm{{m_{\tt H}}}
\def\mbe{{m_{\tt be}}}
\def\msp{{m_\eta}}
\def\rsm{\rho_{\tt sm}}

\def\csm{c_{\tt sm}}
\def\cbe{c_{\tt be}}
\def\li{{\rm Li}}

\def\nsm{n_{\tt H}}
\def\nbe{n_{\tt be}^+}
\def\nbeb{n_{\tt be}^-}
\def\eb{\bar E}

\def\pibf{{\bm\pi}}

\def\eps{\epsilon} 

\def\sp{\eta} 

\def\detail#1{}

\begin{document}

\title{Asymmetric dark matter with a possible Bose-Einstein condensate.}

\author{Shakiba HajiSadeghi}
\author{S. Smolenski}
\author{J. Wudka}

\affiliation{Department of Physics and Astronomy, UC Riverside,\\Riverside, CA 92521-0413, U.S.A}


\begin{abstract}
We investigate the properties of a Bose gas with a conserved charge as a dark matter candidate, taking into account the restrictions imposed by relic abundance, direct and indirect detection limits, big-bang nucleosynthesis and large scale structure formation constraints. We consider both the WIMP-like scenario of dark matter masses $ \gtrsim 1\, \gev $, and the small mass scenario, with masses $ \lesssim 10^{-11}\, \ev $. We determine the conditions for the presence of a Bose-Einstein condensate at early times, and at the present epoch. 
\end{abstract}

\maketitle

\section{Introduction}
\label{sec:introduction}

Understanding the nature of dark matter (DM) remains one of the most pressing contemporary issues in astroparticle physics and cosmology. To date, all DM properties have been inferred from its gravitational effects \cite{Bertone:2004pz}; other probes, such as direct \cite{Sadoulet:1998lyj,Beltrame:2015mpy,Aprile:2017iyp,Serfass:2012gzm} and indirect \cite{Salati:2016zrs,Vitale:2009hr,Bergstrom:1998xh} detection experiments and LHC measurements \cite{PazzinionbehalfoftheATLAS:2017zfa} have produced only limits. These constraints have led to a significant shrinkage of the allowed parameter space in many theoretically favored models \cite{Huang:2017kdh,Olive:2016efh,Jungman:1995df}, and this has spurred interest in alternative models involving dark sectors of varied complexity  \cite{Hwang:2017xmy,Sahoo:2017cqg,Kajiyama:2013rla,Cao:2009uw,Gonzalez-Macias:2016vxy,Drozd:2011aa}. 

A large number of models for DM assume a dark sector that contains one or more dark scalars, which in some cases are the main contributors to the relic abundance required by the CMB experiments \cite{Ade:2015xua}.  Having such scalar relics opens the possibility of such particles undergoing a transition to a Bose-condensed phase; in fact, a variety of models of this kind have been studied in the literature. In some cases the condensate can appear only in the non-relativistic regime, as it happens in axion \cite{Nomura:2008ru,Sikivie:2009qn,Ringwald:2016yge,Sakharov:1994id,Sakharov:1996xg,Khlopov:1999tm,Khlopov:1985jw} and axion-like \cite{Matos:2000ss,Kain:2011pd,UrenaLopez:2008zh,Lundgren:2010sp,Marsh:2010wq,RodriguezMontoya:2010zza,RodriguezMontoya:2011yu,Suarez:2011yf,Harko:2011zt,Harko:2011jy,Matos:2008ag,Chavanis:2011uv,Velten:2011ab,Suarez:2013iw,Magana:2012xe,Aguirre:2015mva,Arbey:2001qi,Hu:2000ke,Sin:1992bg,Lee:1995af,Goodman:2000tg,RindlerDaller:2011kx,Rindler-Daller:2013zxa,Boehmer:2007um,Harko:2011dz} models, where the scalars are assumed to be extremely light. The effects of a Bose-Einstein condensate (BEc)  in such cases have been studied extensively in cosmology  \cite{Matos:2000ss,Kain:2011pd,UrenaLopez:2008zh,Lundgren:2010sp,Marsh:2010wq,RodriguezMontoya:2010zza,RodriguezMontoya:2011yu,Suarez:2011yf,Harko:2011zt,Harko:2011jy,Matos:2008ag,Chavanis:2011uv,Velten:2011ab,Suarez:2013iw,Magana:2012xe,Aguirre:2015mva,Takeshi:2009cy,Fukuyama:2007sx,Fukuyama:2005jq,Dev:2016hxv,Ziaeepour:2010mm} and in astrophysics \cite{Arbey:2001qi,Hu:2000ke,Suarez:2013iw,Sin:1992bg,Lee:1995af,Goodman:2000tg,RindlerDaller:2011kx,Rindler-Daller:2013zxa,Boehmer:2007um,Harko:2011dz}, especially in the context of galactic dynamics, where quantum effects of these very light scalars address the cusp vs. core  \cite{deBlok:2009sp} and ``too big to fail''  \cite{BoylanKolchin:2011de} problems when the scalar mass is $ \sim O(10^{-22})\, \ev $ (though  simulations including both baryonic and Bose-gas components are still lacking). Recently, the authors of Ref.  \cite{Irsic:2017yje} investigated the effects of these light bosons on the Lyman $\alpha$ forest and gave a {\em lower}  bound on the scalar mass $ \gtrsim O(10^{-20})\, \ev $ that excludes the favored mass range, though this result is still being debated   \cite{Zhang:2017chj}.

A prerequisite for the possible appearance of a BEc is the existence of a conserved charge, which is associated with a chemical potential. The simplest model of this type involves a single complex scalar field $ \chi $, and a $\ui$ symmetry,
\beq
\chi \to e^{i\alpha } \chi\,, \quad (\alpha=\mbox{const.})
\label{eq:ui}
\eeq
that  leads to the  required conservation law. Models without an exact conservation law can still exhibit a BEc, but only in the non-relativistic regime, where particle number plays the role of a conserved charge; in these cases the condensate  necessarily disappears as the temperature approaches the particle mass. In contrast, the presence or absence of a condensate in models with a conserved charge is determined by the temperature and density of the gas, in particular, relativistic gases of this sort can condense if the density is sufficiently high.

In this paper we will study several aspects of a dark matter model that obeys \cref{eq:ui} in a flat, homogeneous and isotropic universe. The thermodynamic parameters then will include the corresponding chemical potential ~\footnote{The explicit definition of $ \mu $ is given in \cref{eq:nbe} below; in the non-relativistic regime it is customary to define a shifted quantity $ \mu' = \mu-\,\mbe $ so that condensation corresponds to the condition $ \mu'=0 $.} $ \mu $ assumed to be non-vanishing. The condition $ \mu \not=0 $ presupposes  the presence of a primordial charge whose possible origin we will not discuss in this paper. We will consider two mass regions for the mass $ \mbe $ of the DM particle: {\it(i)} $ \mbe \ge 1\, \gev$ where the behavior in many situations is WIMP-like; and {\it(ii)} $ \mbe \lesssim 2\times 10^{-11} \, \ev $ where the gas can exhibit a condensate at the present epoch.

The model we consider has then the Lagrangian
\beq
\lcal = |\partial \chi|^2  - \mbe^2 |\chi|^2  - \half \lbe |\chi|^4 + \eps |\chi|^2 |\phi|^2 + \lcal\sm\,,
\label{eq:model}
\eeq
where $ \phi$ denotes the SM scalar isodoublet and the last term represents the standard model Lagrangian; all standard model particles are invairaint under \cref{eq:ui}.  We assume throughout that the model is in the perturbative regime and that the BE field does not acquire a \vev. If the Higgs potential takes the form $\lambda\sm (|\phi|-v^2)^2/2 $, we require  {\it(i)} $ \epsilon > - \sqrt{\lbe\, \lambda\sm} $ to ensure (tree-level) stability; {\it(ii)}   $(\mbe/v)^2 > \epsilon $ so that $ \vevof\chi=0,\,\vevof\phi\not=0 $; and {\it(iii)} $ 4\pi \gtrsim \lambda\sm,\,\lbe > 0 $, so that the model remains perturbative. 

This is a simple extension of the usual Higgs-portal models that involve a real scalar field.  Various cosmological aspects of this type of model have been studied   \cite{Matos:2000ss,Kain:2011pd,UrenaLopez:2008zh,Lundgren:2010sp,Marsh:2010wq,RodriguezMontoya:2010zza,RodriguezMontoya:2011yu,Suarez:2011yf,Harko:2011zt,Harko:2011jy,Matos:2008ag,Chavanis:2011uv,Velten:2011ab,Suarez:2013iw,Magana:2012xe,Aguirre:2015mva}, with emphasis on the  low mass regime. Here we will be interested in a much wider range of masses, on the relic abundance and direct detection of dark matter, and in studying the conditions under which a BEc can occur. We  work to $O(\lbe )$: though radiative effects are small in most cases (especially in the non-relativistic regime), they play an important role when obtaining the conditions for the presence of a BEc in the early universe (sec. \ref{sec:cosmology}) and in deriving the restrictions from big-bang nucleosynthesis when the  DM is very light (sect. \ref{sec:BEC}).

In the usual Higgs-portal models \cite{Patt:2006fw,McDonald:1993ex}, for a given choice of DM mass, the relic abundance and direct detection constraints  impose, respectively, lower and upper limits on the DM self coupling constant, and these limits are consistent onlyin a restricted range of masses ($ 55 \, \gev < \mbe < 62 \, \gev$ or $ \mbe > 400\, \gev$) \cite{Athron:2017kgt}; in particular, light masses are excluded. The model \cref{eq:model} sidesteps some of these constraints because  the relic abundance depends on the mass $ \mbe $, the portal coupling $ \epsilon $  and  $ \mu $; the possibility of adjusting the chemical potential relaxes the constraints on the first two parameters (the more severe restrictions found in the simplest Higgs-portal models reappear if one requires $ \mu =0 $).

The BE gas may or may not be in equilibrium with the SM. This is determined by the strength of the coupling $ \epsilon $ in \cref{eq:model} and by the rate of expansion of the universe. As long as the gas and the SM are in equilibrium, they will have the same temperature; when the gas and SM are not in equilibrium they can have different temperatures, but even then the gas will be in equilibrium with {\em itself} and behave as a regular statistical system. In most publications the  relic abundance is calculated using the Boltzmann equation to determine the DM abundance through the decoupling era and into the late universe. We will follow a different approach based on the Kubo formalism \cite{Kubo:1957mj,Grzadkowski:2008xi} that can be used to describe the decoupling of two statistical systems; since the Bose gas remains a statistical system after decoupling such an approach is desirable. For the relic abundance calculation we will use the naive criterion, where decoupling occurs when the interaction rate falls below the Hubble parameter. We do this for simplicity, but also because the presence of a chemical potential allows us to adjust the relic abundance to the experimentally required value, so the full calculation using the kinetics of a Bose gas is not warranted.

The rest of the paper is organized as follows: in the next section we discuss the cosmology of a Bose gas to first order~\footnote{See appendix \ref{sec:app.A} for a summary of the perturbative expansion.} in $ \lbe $ and discuss some aspects of the conditions under which a condensate is present. We next consider relic abundance and the decoupling transition (section \ref{sec:width}) and direct detection (section \ref{sec:cross_section}) in the WIMP regime. We discuss the low-mass scenario in section \ref{sec:BEC}, including constraints from large scale structure formation and big-bang nucleosynthesis. Section \ref{sec:comments} contains parting comments and conclusions, while the appendices involve some formulae used in the text.

\section{Cosmology with a Bose gas}
\label{sec:cosmology}

As mentioned in the introduction, we will consider the behavior of a Bose gas in an expanding universe, including the possibility that a Bose-Einstein condensate (BEc) may be present in some epoch. We will assume that the rate of expansion of the universe is sufficiently slow that the gas will be in local thermodynamic equilibrium~\footnote{This is discussed in detail in \cite{Aguirre:2015mva}.}. To zeroth order in $ \lbe $ (defined in \cref{eq:model}) the thermodynamics quantities correspond to the well-know expressions for an ideal Bose gas \cite{Pathria}. The $O( \lbe) $ can be obtained using standard perturbative methods; we summarize the results in appendix \ref{sec:app.A}. In the calculations below we neglect the $O(\epsilon) $ contributions (cf. \cref{eq:Psmdm}), where  $ \epsilon $ is the portal coupling (cf. \cref{eq:model}) since they are subdominant for the range of parameters being considered in this section: $ \mbe \lesim \msm $ and $ |\epsilon | \lesim \lbe $(see appendix \ref{sec:app.A}).

The occupation numbers for particles and antiparticles are given by
\begin{align}
\nbe  &= \left( e^{(E - \mu)/T}-1 \right)^{-1} = \left( e^{\x(\sqrt{u^2+1}\, - \, \varpi)}-1 \right)^{-1} ; 
\quad \x = \frac \mbe T\,,~ \varpi=\frac\mu\mbe \,.\cr
\nbeb  &= \left( e^{(E + \mu)/T} - 1 \right)^{-1} = \left( e^{\x(\sqrt{u^2+1}\, + \, \varpi)}-1 \right)^{-1} ,
\label{eq:nbe}
\end{align}
where $E = \sqrt{\pp^2+m^2} $ and $ u = |\pp|/\mbe$. 

Defining (see \cref{eq:del.F})
\beq
\delta = \frac{\mu^2  - \mbe^2}\lbe\,,~~ \fBB = 2\int \frac{d^3\pp}{(2\pi)^3 2E}  \left[ \nbe + \nbeb \right]_{\mu = \mbe} \,,
\eeq
the phase transition line is given by
\beq
\delta = \fBB\,.
\eeq
A condensate will not form if $\mu^2 < \mbe^2 + \lbe \fBB$; when $ \lbe =0 $ this reduces to the well-known result that a condensate is present only if $ |\mu| = \mbe $.

The conserved charge associated with the symmetry of \cref{eq:ui} is given by
\begin{align}
\ch &= \ch\up c + \ch\up e \cr
  &= \ch\up c + \mbe^3 \nu\be \,;\quad
\nu\be =  \int_0^\infty \frac{du \, u^2}{2 \pi^2} (\nbe - \nbeb) + O(\lbe) \,,
\label{eq:nu}
\end{align}
where $ \ch\up{e,c}$ are the charge densities in the excited states and in the condensate (if present). Without loss of generality we will assume  $\ch\up c \ge 0$; if there is a condensate then $ \mu > 0 $.

The entropy and energy densities for the Bose gas are given by
\begin{alignat}{2}
s\be &= \mbe^3 \sigma\be\,; 
&&  \sigma\be =   \int_0^\infty \frac{du \, u^2}{2 \pi^2} \sum_{n=n\be^\pm} \left[ (1+ n) \ln(1 +n)  - n \ln n \right] + O(\lbe) \,,\cr
\rho\be &= \ch \mu + T\sbe - P\be \cr
&= \mbe \ch\up c + \mbe^4 r\be\,; 
&& \quad r\be =   \int_0^\infty \frac{du \, u^2}{2 \pi^2} \sqrt{u^2+1} ( \nbe + \nbeb) + O(\lbe) \,.
\label{eq:rbe-sbe}
\end{alignat}

The $O(\lbe)$ corrections are given in \cref{eq:qs_nobec} and \cref{eq:qs_bec}, and though we will use them in the calculations below, they are not displayed so as not to clutter the above expressions.

The Standard Model energy and entropy densities are approximately given by \cite{Kolb:1990vq}
\beq
\rsm =  \frac{\pi^2}{30}T^4 \gs(T) \,, \qquad  \ssm =  \frac{2\pi^2}{45} T^3 \gss(T)\,,
\label{eq:rsm-ssm}
\eeq
where
\begin{align}
\gs(T) &\simeq \sum_{\rm bosons} g_i \left( \frac{T_i}T \right)^4 \theta(T - m_i ) + \frac78 \sum_{\rm fermions} g_i \left( \frac{T_i}T \right)^4 \theta(T - m_i ) \,,\cr
\gss(T) &\simeq \sum_{\rm bosons} g_i \left( \frac{T_i}T \right)^3 \theta(T - m_i ) + \frac78 \sum_{\rm fermions} g_i \left( \frac{T_i}T \right)^3 \theta(T - m_i ) \,,
\label{eq:gs}
\end{align}
where $g_i$ denotes the number of internal degrees of freedom, and $T_i$ the temperature for each particle; we assumed a zero chemical potential for the SM particles.

In the discussion below we repeatedly use the fact that when the SM and Bose gas are in equilibrium with each other the ratio $ \ch/s\tot $ is conserved, where $ s\tot = \sbe + \ssm$ is the total entropy. When the SM and Bose gas are not in equilibrium with each other the ratios $ \ch/\ssm $ and $ \sbe/\ssm $ are separately conserved (in this case $ \ch/s\tot $ is also conserved, but it is not independent of these quantities).

\section{The Bose-Einstein condensate}
\label{sect:condensate}
As noted above, whether the SM and gas are in equilibrium with each other or not,  the ratio $Y$
\beq
Y = \frac{\ch}{s\tot}
\eeq
is conserved (though the $(e)$ and $(c)$ contributions in general are not). A condensate will  be present whenever the total charge cannot be accommodated in the excited states, that is, when $ Y > Y\up e $:
\beq
\ch\up c \not=0 \quad {\rm if} \quad Y >   Y\up e = \left. \frac{ \nu\be}{ \sigma\be + \ssm/\mbe^3 }\right|_{\delta=\fBB}\,.
\label{eq:bec.forms}
\eeq
Now, since $ \ssm > 0 $, we have the following inequality:
\beq
 Y\up e  < \left. \frac{ \nu\be}{ \sigma\be}\right|_{\delta=\fBB} <  \left. \frac{ \nu\be}{ \sigma\be}\right|_{\delta=\fBB\,, T \to 0 } = \frac{ \zeta_{3/2}}{ (5/2)\zeta_{5/2}} \simeq 0.78\,.
\label{eq:Y.bound}  
\eeq
Therefore, a condensate will be always present if $ Y > 0.78 $. 

\begin{figure}[ht]
$$
\includegraphics[width=4in]{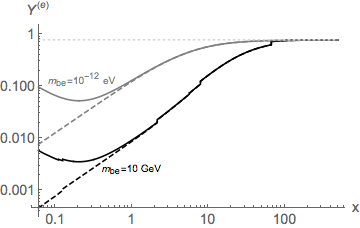} 
$$
\caption{Plot of the Bose charge in the excited states per entropy when $ \lbe=0.5$ (solid curves) and $ \lbe = 0$ (dashed curves) and for two mass values
and $ \mbe = 10\, \gev $ (black curves) $ \mbe = 10^{-12}\, \ev $ (gray curves); the dotted line corresponds to the bound in \cref{eq:Y.bound}. For illustration purposes we assumed the Bose gas and the SM have the same temperature. The discontinuities are caused by the step functions in \cref{eq:gs} and $x = \mbe/T $ (cf. \cref{eq:nbe})}
\label{fig:Ye}
\end{figure}

The behavior of $ Y\up e $ for various choices of $ \mbe$ and $ \lbe $ is given in figure~\ref{fig:Ye}. For large temperatures~\footnote{The Bose gas entropy and charge are not exponentially suppressed as $ T \to 0 $ when $ |\mu| = \mbe + O(\lbe)$.} and $ \lbe=0$, $ \nu\be/\sbe \sim 1/T $ (cf. \cref{eq:rel.rns}) since the leading particle and antiparticle contributions to $ \nu\be$ in \cref{eq:nu} cancel; it follows that $ Y\up e(\lbe=0) \to 0 $ as $ T \to \infty $, in particular, in an ideal gas a condensate would always be present at sufficiently high temperatures~\footnote{This holds whether the SM and Bose gas are in equilibrium or not.} \cite{Haber:1981ts}. This behavior changes when $ \lbe \not=0 $: $Y\up e$ has an $ \mbe$-dependent minimum~\footnote{For a discussion of the validity of our expressions in this region see appendix \ref{sec:app.A}.}, so that a self-interacting BE gas with a sufficiently small $Y$ will never condense. If the behavior of $Y\up e$ to $O(\lbe)$ is indicative of the exact result, then $ Y\up e $ diverges as $ x \to 0 $ and the condensate will disappear for sufficiently high temperatures, this is discussed further in appendix \ref{sec:app.A}.

To clarify this behavior note that in an expanding universe both the (co-moving) volume and temperature change with $a$, the distance scale in the Robertson-Walker metric, with
$ T \to \infty $ as  $ a\to 0 $: a contracting co-moving volume accompanies an increasing temperature. There are then two competing effects on the Bose gas: the reduction of volume favors the formation of the condensate, while the increase in temperature tends to destroy it; the above results indicate that when $ \lbe =0 $ the volume effect dominates. When $ \lbe \not=0 $ a third effects comes into play: the repulsive force generated by the Bose gas self-interactions, which gives rise to the non-monotonic behavior of $Y\up e$.

Because of the exact $ \ui $ symmetry of the dark sector, the presence of this condensate does not require the gas to be non-relativistic (in which case $\ch$ is equals particle number). We will see later (see section \ref{sec:BEC}) that experimental constraints allow for the condensate to persist to the present day only if $ \mbe $ is in the pico-\ev\ range; for WIMP scenarios ($\mbe \gtrsim 1 \gev $) the condensate disappears already in the very early universe.

\subsection{Conditions for a BEc at decoupling}
We will show below that for WIMP-like masses ($ \mbe \gtrsim 1 \, \gev$) the gas and SM decouple at a temperature $T_d$, at which point the gas will be non-relativistic; it then follows that it will also be non-relativistic at present. In the non-relativistic limit the $ O(\lbe)$ corrections to the  expressions below can be ignored since they are smaller than the $O(T/\mbe)$ relativistic corrections (see  \cref{eq:Pnr} and surrounding discussion in appendix \ref{sec:app.A}). Then
\beq
\frac\ch\ssm \simeq \inv\mbe \frac{\rho_{\rm DM}}\ssm = \frac{0.4\, \ev}\mbe\qquad ( T< T_d)\,,
\label{eq:rel.ab}
\eeq
where we used the known value of the SM entropy now, and the fact that for a non-relativistic gas $ \rho_{\rm DM}=\mbe \ch $; as noted in section \ref{sec:cosmology}, the \lhs of \cref{eq:rel.ab} is conserved below $T_d$.  

This can be used to determine whether a BEc would have been present when $T=T_d$: a condensate is present if
\beq
\frac{\ch(T_d)}{( \mbe T_d)^{3/2}}   > \frac{\zeta_{3/2} }{(2\pi)^{3/2}} \simeq 0.166\,.
\label{eq:cond}
\eeq
using \cref{eq:rel.ab} to eliminate $ \ch(T_d) $ and \cref{eq:rsm-ssm} for the SM entropy, this implies
\beq
\frac{T_d^{3/2}}{ \mbe^{5/2}} \gss(T_d) > \inv{1.06\, \ev} \,.
\eeq
whence, since for a non-relativistic gas $ \mbe > T_d$, and since $ \gss < 106.75 $, we find (using $3\sigma$ errors)
\beq
\mbe < 1.3 \kev\,.
\eeq
A condensate can occur at decoupling only for light Bose particles, which can be difficult to accommodate phenomenologically (cf. section \ref{sec:BEC}).

\subsection{Conditions for a BEc to exist at present}
Before proceeding with the calculation of the cross section relevant for direct detection, we study the possibility that the Bose gas supports a condensate at present. To this end we note first that a non-relativistic Bose gas will have a condensate provided $ \ch ( \mbe T)^{-3/2} > \zeta_{3/2} (2\pi)^{-3/2}$, see \cref{eq:cond}; denoting the current gas temperature by $T_{\tt now}$ it follows that a condensate will be currently present if
\beq
\left(\frac{0.0215 \, \ev}\mbe\right)^{5/3}\, ^o{\rm K} > T_{\tt now}\,.
\label{eq:BEC.now}
\eeq
We now use the fact that the conservation of $ \sbe/\ssm $ allows us to obtain a relation between $ T_{\tt now} $ and the decoupling temperature $T_d $. Noting that the gas is non relativistic at $T_d $, and that a condensate at $ T_{\tt now} $ implies a condensate was also present at $ T_d $ (see section \ref{sec:cosmology}), we find
\beq
 \frac{4.3 \, ^o{\rm K}}{\gss(T_d)^{1/3}} = \sqrt{T_d T_{\tt now}}\,,
 \label{eq:s.ratio}
\eeq
where we used \cref{eq:rbe-sbe} and \cref{eq:rsm-ssm}. Combining this with \cref{eq:BEC.now} and using \cref{eq:nrel.rns},
\beq
[\gss(T_d)]^{2/3}\, T_d  \gesim \left(\frac\mbe{0.009 \, \ev}\right)^{5/3} \, ^o{\rm K}\,,
\eeq
and since $ \mbe > T_d$, this gives
\beq
9.5\,\gss(T_d)\, \ev \gesim \mbe \then 88\, \ev > \mbe\,.
\eeq
It follows that  a WIMP-like Bose gas will not exhibit a condensate at the present era~\footnote{Since the gas is again non-relativistic  the $O(\lbe) $ corrections to the above expressions can be ignored; see \cref{eq:Pnr}.} (nonetheless, for completeness we include in Appendix \ref{sec:app.B} the expressions for the cross section when a condensate does occur). The case of a light Bose gas with a condensate will be considered in section \ref{sec:BEC}.

\subsection{The BEc transition temperature:}
For WIMP-like masses we will show (section \ref{sec:width}) that the SM and Bose gas will be in equilibrium down to a decoupling temperature $ T_d$. Below $T_d$ the ratios $ \ch/\ssm$ and $ \sbe/\ssm $ will be separately conserved, above $T_d$ only $ \ch/s_{\tt tot} $ is conserved. We will also show that in this case the gas was non-relativistic at $T=T_d$ and that the relic abundance constraint reduces to the simple relation $ \ch= 0.4\, \ev (\ssm/\mbe) $ (cf. 
\cref{eq:rel.ab}). Combining these results we find that the temperature $T_{\tt BEC}$ at which the condensate forms (the same for the gas and SM since $ T_{\tt BEC}> T_d$) is given by
\beq
 \left[ 2+ \gss(T_{\tt BEC}) \right] T_{\tt BEC} = \frac{15}{2\pi^2} \left[ \frac52 - \ln z_d + \frac\mbe{0.4\, \ev} \right] \mbe \then  T_{\tt BEC} \simeq \mbe^2 \frac{ 1.9 \, \ev^{-1} }{\gss( T_{\tt BEC}) + 2 }\,,
 \eeq
 where~\footnote{It follows from \cref{eq:nrel.rns} and the conservation laws that $z$ is constant below $T_d$ for a non-relativistic gas without a condensate.} $ z = \exp[(\varpi-1)\x] $, and we used the fact that $ |\ln z |\ll \mbe/(0.4 \ev)$ for all cases being considered. As noted previously, the $O(\lbe)$ corrections can be ignored in these calculations; the subscript $d$ denotes a quantity at decoupling.
 
For example, $  T_{\tt BEC} \sim 10^7\, \gev $ if $ \mbe \sim 1\, \gev $ and $\gs( T_{\tt BEC}) \sim 100 $  (though, of course, the number of relativistic degrees of freedom at these high temperatures may be much higher); while $  T_{\tt BEC} \sim 1.75\, \tev $ if $\gs( T_{\tt BEC}) =106.75$ and $ \mbe \sim 10 \mev$. It is worth noting that for the WIMP-like scenario, the condensate, should it form, would hold a small fraction of the total energy density of the gas: using \cref{eq:rel.rns} and \cref{eq:nrel.rns} and the above conservation laws we find, 
\bal
\left. \frac{\mbe \ch}{\rho\be} \right|_{T>T_{\tt BEC}} &= \frac{ 2+ \gss(T) - (5/\pi^2)A\, \x }{ 2+ \gss(T) + A \, \x^{-1}} \,, \quad A = \frac32 \left[ \frac52 - \ln z+ \frac\mbe{0.4\, \ev} \right]\simeq \frac\mbe{0.27\, \ev} \\[5pt]
&\simeq (0.27\, \ev) \frac{ 2+ \gss(T)}T\, \quad \mbox{for}~x \ll 0.4\, \ev/\mbe\,.
\end{align}
So in the early universe $ Y\up e \to 0 $ but $ \rho\up e\be /\rho\be \to 1 $:  the charge resides mainly in the condensate,  but the energy is carried mainly by the excited states.

For an ultra-light DM ($ \mbe \sim 10^{-12} \, \ev$) the situation is completely different. We discuss this in section \ref{sec:BEC}.

\section{Relic abundance}
\label{sec:width}

In obtaining the relic abundance we will follow an approximate method that will not involve solving the Boltzmann equation. Instead we imagine the Bose gas and the SM to be in equilibrium at some early time and describe their decoupling using the Kubo formalism \cite{Kubo:1957mj}. As we see below, the BE gas will be non-relativistic, so that in this section the $ O(\lbe) $ corrections can be ignored (see appendix \ref{sec:app.A}).

The total Hamiltonian for the system is  of the form
\beq
H = H\sm + H\be - H'\,, \quad H' = - \eps \int d^3x\, \ocal\sm \ocal\be\,,
\eeq
where $ \ocal\sm = |\phi|^2\,~ \ocal\be =|\chi|^2 $  and $ \eps $ is defined in \cref{eq:model}. Using the  same arguments as in \cite{Grzadkowski:2008xi}, we find that the temperature difference (and hence a lack of equilibrium) between the SM and Bose gas obeys
\beq
\dot \vartheta + 4 \hBB \vartheta = - \Gamma \vartheta\,; \quad \vartheta = T\be - T\sm\,,
\eeq
where $ \hBB$ is the Hubble parameter. This expression is valid when $ \vartheta \ll T\be,{}\sm $, so the width $ \Gamma$ can be evaluated at the (almost) common temperature $T$. We use this expression to define the temperature $ T_d$ at which the SM and Bose gas decouple by the standard condition \cite{Kolb:1990vq}
\beq
T = T_d \then \Gamma = \hBB \,.
\label{eq:Td} 
\eeq

Explicitly we have \cite{Grzadkowski:2008xi}, 
\beq
\Gamma = \left( \inv{\cbe} + \inv{\csm} \right) \frac{\eps^2 G}T\,,
\label{eq:Gamma}
\eeq
where $ \csm,~\cbe $ denote the heat capacities per unit volume, $T$ the common temperature, and
\beq
G = \int_0^\beta ds \int_0^\infty dt \int d^3\xx \vevof{\ocal_{\rm BE}(-is,\xx) \dot\ocal_{\rm BE}(t,\bfz)} \vevof{\ocal_{\rm SM}(-is,\xx) \dot\ocal_{\rm SM}(t,\bfz)} \,.
\label{eq:G}
\eeq

The  heat capacities are given by
\bal
\csm &= \frac{4\pi^2}{30} T^3 \gss\,;\mcr
\cbe &= \left( \frac{\mbe T}{2\pi} \right)^{3/2} \times \left\{
\begin{array}{ll} 
(15/4) \li_{5/2}(1) & \mbox{(BEc)}\,,\cr 
(15/4) \li_{5/2}(z) -(9/4)[\li_{3/2}(z)]^2/\li_{1/2}(z)& \mbox{(no~BEc)}\,,
\end{array} 
\right.
\end{align}
where $ \li$ denotes the Poly-logarithmic function, and $ z = \exp[(\mu - \mbe)/T] $.

\subsection{Evaluation of $G$}
In the presence of a condensate we follow \cite{Kapusta:1981aa} and write $ \chi = [(A_1+\bec) + i A_2]/\sqrt{2} $, where $A_{1,2}$ denote the fields and $\bec$ the condensate amplitude. We also assume that decoupling occurs below the electroweak phase transition so that $ |\phi|^2 = ( v + h)^2/2 $, where $v$ is the SM \vev, and $h$ the Higgs field. Substituting in \cref{eq:G} we find, after an appropriate renormalization,
\beq
G_{\hbox{$\scriptstyle \tt BEc$}} = \left[v^2 \bec^2 G_{2-2} + \inv4 \bec^2 G_{2-4} + \inv4 v^2 G_{4-2} + \inv{16}G_{4-4}\right]_{\mu = \mbe }\,,
\eeq
where
\bal
G_{2-2} &= \int_0^\beta ds \int_0^\infty dt \int d^3\xx  \vevof{A_1(-i s,\xx) \frac{d A_1(t,\bfz)}{dt}} \vevof{h(-i s,\xx) \frac{d h(t,\bfz)}{dt}} \,, \mcr
G_{2-4} &= \int_0^\beta ds \int_0^\infty dt \int d^3\xx  \vevof{A_1(-i s,\xx) \frac{d A_1(t,\bfz)}{dt}} \vevof{h^2(-i s,\xx) \frac{d h^2(t,\bfz)}{dt}} \,, \mcr
G_{4-2} &= \int_0^\beta ds \int_0^\infty dt \int d^3\xx  \vevof{\AA^2(-i s,\xx) \frac{d \AA^2(t,\bfz)}{dt} } \vevof{h(-i s,\xx) \frac{d h(t,\bfz)}{dt}} \,, \mcr
G_{4-4} &= \int_0^\beta ds \int_0^\infty dt \int d^3\xx  \vevof{\AA^2(-i s,\xx) \frac{d \AA^2(t,\bfz)}{dt}} \vevof{h^2(-i s,\xx) \frac{d h^2(t,\bfz)}{dt}}\,.
\end{align}

In the absence of a condensate we have
\beq
G_{\hbox{\sout{$\scriptstyle \tt BEc$}}} =  \inv4 v^2 G_{4-2} + \inv{16}G_{4-4}\,,
\eeq
($G_{\hbox{\sout{$\scriptstyle \tt BEc$}}} $ denotes the expression for $G$ in the absence of a condensate) evaluated at a chemical potential $ |\mu|  < \mbe $.

We evaluate the $ G_{n-m}$ using the standard Feynman rules for the real-time formalism of finite-temperature field theory (see for example  \cite{Bellac:2011kqa}) and the propagators derived in \cref{sec:app.A}. The calculation is straightforward but tedious; to simplify the expressions we use the following shortcuts:
\beq
\begin{array}{llll}
E = E_\kk\,, &\quad  E' = E_{\kk'}\,, &\quad  \eb = \eb_\qq\,, & \quad \eb' = \eb_{\qq'}\,, \cr
\nsm = \nsm(E_\kk)\,,&\quad \nsm'  = \nsm(E_{\kk'})\,, &\quad n\be^\pm=n\be^\pm(\eb_\qq)\,,&\quad n\be^\pm{}'=n\be^\pm(\eb_{\qq'})\,,
\label{eq:shortcuts}
\end{array}
\eeq
and
\beq
\dt k = \frac{d^3\kk}{2 E_\kk (2\pi)^3}\,,\qquad \dt q = \frac{d^3\qq}{2 \eb_\qq (2\pi)^3}\,;
\eeq
where 
\beq
E_\kk = \sqrt{\msm^2 + \kk^2}\,, \qquad  \eb_\qq = \sqrt{\mbe^2 + \qq^2} \,; \qquad n\be\up\pm(\eb) = \left[e^{\beta (\eb\mp \mu)} - 1 \right]^{-1}\,,
\label{eq:E.n}
\eeq
and $ \msm $ denotes the Higgs mass.

Then the $G_{n-m}$ (for arbitrary $ \mu $) are given by
\bit
\item{$G_{4-4}$}
\bal
G_{4-4} &= 16 \pi \beta \int \dt k \dt{k'} \dt q \dt{q'} (2\pi)^3 \, \delta\up3(\kk+\kk'+\qq+\qq') \gcal_{4-4} \,; \mcr 
\gcal_{4-4} &=  \half   (1+\nsm) (1+\nsm') \nbe  \nbeb{}' \, \delta(E + E' - \eb - \eb') \, (E+ E')^2\mcr
& \quad + \half  (1+\nbe) (1+\nbeb{}') \nsm  \nsm' \, \delta(E + E' - \eb - \eb') \, (E+ E')^2\mcr
& \quad + (1+\nsm) (1+\nbe) \nsm' \nbe{}' \, \delta(E + \eb - E' - \eb') \, (E-E')^2 \mcr
& \quad + (1+\nsm) (1+\nbeb) \nsm' \nbeb{}' \, \delta(E + \eb - E' - \eb') \, (E-E')^2\,,
\label{eq:g44}
\end{align}
where the 4 terms represent the processes 
$ h h \leftrightarrow \chi \chi^\dagger$, 
$ h \chi \to h \chi$ and
$ h \chi^\dagger \to h \chi^\dagger$ respectively; the factors of 1/2 are due to Bose statistics.

\item {$ G_{2-4}$}
\bal
G_{2-4} &= 2\pi \beta \int \dtk \dtk' \dtq (2\pi)^3 \delta\up3(\kk+\kk'+\qq) \gcal_{2-4} \,;\mcr
\gcal_{2-4} &= 
  \half  (1+\nsm)   ~ (1+\nsm') \nbeb  \, \delta(E + E' - \eb - \mbe)  \, (E+ E')^2\mcr
& \quad+ \half (1+\nbeb)  \nsm \nsm'  \, \delta(E + E' - \eb - \mbe) \, (E+ E')^2\mcr
& \quad+        (1+\nsm)   \nsm'  \nbe   \, \delta(E + \mbe - E' - \eb)  \, (E-E')^2 \mcr
& \quad+        (1+\nsm)    (1+\nbe)    \nsm'  \, \delta(E + \eb - E' - \mbe) \, (E-E')^2\,,
\label{eq:g24}
\end{align}
these 4 terms represent the processes 
$ h h \leftrightarrow \bec \chi^\dagger$ and
$ h \bec \leftrightarrow h \chi$, where $ \bec$ corresponds to a particle in the condensate (mass $ \mbe$ and zero momentum); the factors of 1/2 are due to Bose statistics.

\item{$ G_{4-2}$}
\bal
G_{2-4} &= 4\pi \beta \int \dtk \dtq \dtq' (2\pi)^3 \delta\up3(\kk+\qq+\qq')  \gcal_{4-2} \,;\mcr
\gcal_{4-2} &=  \left[ (1+\nbe)(1+\nbeb{}') \nsm +  (1+\nsm) ~ \nbe ~ \nbeb{}' \right] E^2 \delta (\eb + \eb' - E)\,,
\label{eq:g42}
\end{align}
these 2 terms represent the processes $ h \leftrightarrow \chi \chi^\dagger$.

\item {$ G_{2-2}$}
\bal
G_{2-2} &= \half \pi \beta \int \dtk \dtq  (2\pi)^3 \delta\up3(\kk+\qq) \gcal_{2-2} \,;\mcr
\gcal_{2-2} &= \left[ (1+\nsm)\nbeb + (1+\nbeb)\nsm(E)   \right] E^2 \delta (E - \mbe - \eb) \,,
\label{eq:g22}
\end{align}
these 2 terms represent the processes $ h \leftrightarrow \bec \chi^\dagger$.
\eit

In the non-relativistic limit, where $ \mbe,\,\msm \gg T$ we find~\footnote{$G_{2-2,\,2-4}$ contribute only when there is  condensate, so we evaluate then them only for $ \mu = \mbe$; the expressions for $G_{4-2,\,4-4}$ are valid for all $ \mu $.}
\begin{align}
\left. G\up{\rm NR}_{2-2}\right|_{\mu = \mbe} &\simeq \frac{   \msm }{r \sqrt{(2\pi)^3 \x}}   \frac{ 2\eg\, e^{ - 2 \x}}{\eg^2+ (r^2-4)^2} \,;\mcr 
\left. G\up{\rm NR}_{2-4}\right|_{\mu = \mbe}  &\simeq  \left(\frac\msm{ 2\pi r \x}\right)^3  \left[2 r^2 \x^2 \rho  K_1(\rho) +  \zeta_3\left( \frac{(r+1)^2}{4r} \right) \right] e^{- r \x}\,;  \mcr
G\up{\rm NR}_{4-2} &\simeq  \left( \frac\msm{2\pi} \right)^3  \frac4{\x^2 r^3}\left[ e^{- r\x}\sqrt{\pi \left(\frac{r \x}2 \right)^3 \left(\frac{r^2}4-1 \right)}  \, \theta(r - 2 ) + \frac{\li_{3/2}(z)}z \frac{2\eg\, e^{-2\x}}{\eg^2 + (r^2-4)^2}\right] \,;
 \mcr
G\up{\rm NR}_{4-4}&\simeq \inv{16}\frac{\msm^5}{r^3 (1+r)^{7/2}}  \left( \frac{2}{\pi \x} \right)^{9/2} e^{ - r\x}\left( z+ \inv z  e^{ - 2\x}  \right) \,,
\label{eq:NRG}
\end{align}
where $K_1,\,\zeta_3$ and $ \li$ denote the usual Bessel, zeta and Poly-logarithmic functions, and we defined
\beq
 r = \frac\msm\mbe, \quad \rho = \frac{4 r |r-1| \x}{\sqrt{2(r^2+1)}}, \quad \eg = r^2 \frac{\Gamma\sm}{\msm}\,, \quad z = e^{\beta(\mu - \mbe)}\,.
\label{eq:r.def}
\eeq
 
Before continuing it is worth pointing out a slight difference between the expression for $ \Gamma $ derived from \cref{eq:G} and \cref{eq:Gamma}, and the corresponding expression usually found in the literature (see {\it e.g.} \cite{Kolb:1990vq}): \cref{eq:Gamma} describes the energy transfer between the SM and the Bose gas, which leads to the $(E \pm E')^2$ factors in \crefrange{eq:g44}{eq:g22}. As a result $ \Gamma $ in \cref{eq:Gamma} has a factor $ \sim ({\rm mass}/T)^2 $ compared to the usual expressions, which determine the change in the DM particle number. Because of this the decoupling temperature obtained from \cref{eq:Td} will be somewhat higher than usual; this difference, however, is not significant given that the criterion \cref{eq:Td} itself is not sharply defined.

\subsection{The decoupling temperature}

For a non-relativistic at $ T = T_d $, we have from \cref{eq:rel.ab} 
\beq
\frac{0.4\, \ev}\mbe \ssm(T_d) \simeq 2 \left( \frac{\mbe T_d}{2\pi} \right)^{3/2}  \cosh( \mu/T_d) \,  e^{ - \mbe /T_d }\,.
\label{eq:el.mu}
\eeq
We will use this expression to eliminate $ \mu $ in \cref{eq:Td}; in doing this we  implement the requirement that the Bose gas generates the correct DM relic abundance~\footnote{This calculation can yield $ |\mu| > \mbe $ for some choice of $ \mbe $ and $T_d $, this only means that such masses and temperatures are excluded by the relic abundance constraint.}

\begin{figure}[h]
$$
\includegraphics[height=2.5in]{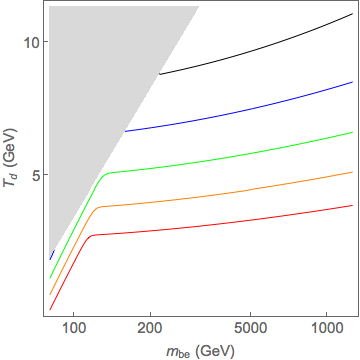} 
\quad
\includegraphics[height=2.5in]{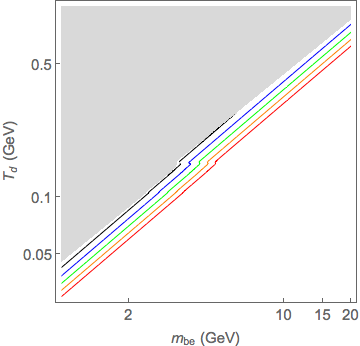}
$$
$$
\includegraphics[height=2.5in]{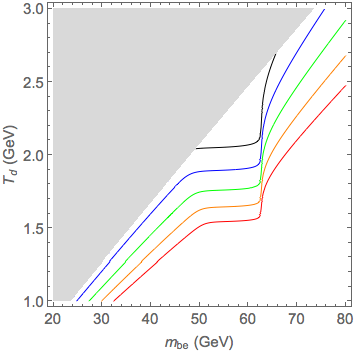}
$$
\caption{Values of $T_d$ satisfying the decoupling condition \cref{eq:Td} as a function of $\mbe $ for $ \epsilon=0.001,\,0.01,\,0.1,\,1,\,10$ (bottom to top curves) and for low and high values of $\mbe$ (top left and top right graphs, respectively), and in the resonance region (bottom graph). The trough at $ \mbe\simeq 62.5 \, \gev $ corresponds to the effects of the Higgs resonance. The shaded region is excluded by the relic abundance constraint.}
\label{fig:m-T}
\end{figure}

Using then \cref{eq:el.mu} to eliminate $ \mu $, the condition $ \Gamma = \hBB$ in \cref{eq:Td} provides a relation between $ T_d,\,\mbe$ and $ \eps $, which we plot in Fig. \ref{fig:m-T}. The resonance effects are broadened  below $ \msm/2 $ due to the effects of the non-resonant term in $ G_{4-2} $ that are proportional to $ \theta( r-2) $. The rapid change in curvature observed for $ \mbe \sim 100\, \gev $ is produced by $ G_{4-4} $, which dominates $ \Gamma $ for large masses. We also see that, for the range of couplings being considered, $ T_d \lesssim \mbe/10$ so that the gas is non-relativistic at decoupling, as was assumed above. 

\section{Direct detection}
\label{sec:cross_section}

We first calculate the cross section for the process $\sp\chi\rightarrow\sp\chi$, where $ \eta $ denotes a neutral scalar coupled to the Bose gas via an interaction
\beq
\lcal_{\sp-\chi} = \half g \sp^2 |\chi|^2\,.
\label{eq:L_sp-chi}
\eeq
The interesting case of nucleon scattering will reduce to the expressions obtained for $ \eta $ in the non-relativistic limit (for an appropriate choice of $ g $), except for a spin multiplicity factor. 

The transition probability is given by
\beq
W_{i\rightarrow f}=|_{\tt out}\vevof{f\,|i}_{\tt in}|^{2}\,,
\eeq
where the initial state consists of an $\sp$ particle with momentum \pp\ and the Bose gas in state {\tt I}: $\ket i_{\tt in}=a_{\mathbf{p}}^{\tt in\,\dagger}\ket{0;{\tt I}}$ (where $0$ denotes the perturbative vacuum for the $\sp $); the final state has an $ \sp $ of momentum \qq\ and the Bose gas in a state {\tt F}: $\ket f_{\tt out} = a_{\mathbf{q}}^{\tt out\,\dagger}\ket{ 0;{\tt F}}$.  We require  $\pp \neq\qq$, since we are looking
for non-trivial interactions. 

Using the standard LSZ reduction formula we find~\footnote{We work to $O(g)$ and assume a non-relativistic gas, so the $ O(\lbe) $ corrections can be ignored.} 
\bal
_{\tt out}\vevof{f\,|i}_{\tt in} &=\vevof{0;{\tt F} \left| \Theta_{\pp,\qq} \right|0;{\tt I}}\,, \mcr
\Theta_{\pp,\qq} &= -\int d^{4}x\,d^{4}x'e^{-ip\cdot x+iq\cdot x'}(\Box_{x}+m^{2})(\Box_{x'}+m^{2})\, \tBB\left[\sp(x)\,\sp(x')\right]\,,
\end{align}
where $ \tBB$ is the time-ordering operator and we ignored a wave-function renormalization factor (we will be working to lowest non-trivial order, where this factor is one). In order to obtain the cross section, we sum over the final gas states ({\tt F}) and thermally average over initial gas states ({\tt I}); this gives
\bal
&\vevof{W_{i\rightarrow f}}_\beta= \int d^{4}x\,d^{4}x'\,d^{4}y\,d^{4}y'e^{i(p\cdot y-q\cdot y' - p\cdot x + q\cdot x')} (\Box_{x}+m^{2})(\Box_{x'}+m^{2})\cr
 & \qquad \times (\Box_{y}+m^{2})(\Box_{y'}+m^{2}) \vevof{ \tBB\left[\sp(x^{0}-i\beta,\mathbf{x})\sp(x'^{0}-i\beta,\mathbf{x'})\sp(y^{0},\mathbf{y})\sp(y'^{0},\mathbf{y'})\right]}_{\beta}\,,
 \label{eq:Wfi}
\end{align}
where $ \vevof\dots_\beta$ indicates a thermal average at temperature $ 1/\beta $. $\vevof{W_{i\rightarrow f}}_\beta$ can be evaluated using standard techniques of the real-time formulation of finite-temperature field theory~\footnote{In particular, under $\tBB$, the complex times in \cref{eq:Wfi} are later than the real ones.} \cite{Bellac:2011kqa}, while the optical theorem relates this quantity to the desired cross section:
\beq
\sigma = \inv{2 \ch |\pp|} \left( \inv{\vcal} \int' \frac{ d^3 \qq}{2 E_\qq\, (2\pi)^3}\, \vevof{W_{i\rightarrow f}}_\beta \right)\,,
\label{eq:DD-sigma}
\eeq
where $ E_\qq$ is the energy of the outgoing $\sp $, $\ch$ the number density of Bose gas particles, and $ \vcal $ denotes the volume of space-time; the prime indicates that the region $ \pp \simeq \qq $ is to be excluded.

To lowest order in $ g $ (see \cref{eq:L_sp-chi}) we have
\beq
\vevof{W_{i\rightarrow f}}_\beta= g^2 \int \frac{d^4k}{(2\pi)^4} \left. \left[D^<(k+P) \right]_{ij} \left[D^>(k)\right]_{ij} \right|_{\bec=0}\,; \quad P = p-q\,,
\eeq
where the propagators are given in \cref{eq:D.f} and \cref{eq:rhobe}, and $\bec=0$ implements the absence of a condensate. The evaluation of this expression is straightforward, we find
\bal
\vevof{W_{i\rightarrow f}}_\beta
&= \frac{g^2 T f(-P_0)}{2\pi|\PP|} \ln \left| \frac{1 + \nbe (E_-)}{1+\nbe(E_+)} \frac{1+\nbeb(E_-)}{1+\nbeb(E_+)} \right| \,,\mcr
&\simeq\frac{g^{2}}{4\pi|\PP|\beta}e^{-\beta E_-}\, \cosh(\beta\mu)\,; \quad E_\pm=\half \left[ |\PP|\sqrt{1-\frac{4\mbe^2}{P^2}} \mp P_0 \right]\,,
\label{eq:Wif.no.BEC}
\end{align}
where $ n\be\up\pm$ are defined in \cref{eq:shortcuts}, and $f$ in \cref{eq:D.f}; the second expression is valid in the non-relativistic limit. Substituting this into \cref{eq:DD-sigma} gives
\bal
\sigma &= \left[ \inv{\sqrt{\pi}\, u} e^{-u^2} + \left( 1 + \inv{2 u^2}\right) {\rm Erf}(u)  \right]\sigma_0 \,; \quad u = \frac{|\pp|}\msm\sqrt{ \frac\mbe{2 T}} \,, \mcr
&=\left[1 + \inv{2 u^2} + O\left( u^{-5} e^{-u^2} \right) \right] \sigma_0 \,, \quad (u\to \infty )
\end{align}
where $ \sigma_0 $ is the $T=0$ non-relativistic cross section, and in \cref{eq:DD-sigma} we used
\beq
n = 2 \left( \frac{\mbe T}{2\pi} \right)^{3/2} e^{ - \beta \mbe} \, \cosh(\beta\mu)\,.
\eeq

The above expression for $\vevof{W_{i\rightarrow f}}_\beta$ holds also for non-relativistic nucleons, except for a  factor of $2 m_N^2$, where $ m_N$ is the nucleon mass. Also, since for the direct-detection reactions the momentum transfer for this process is very small, the coupling $ g $ will be given by
\beq
g \to \frac{\epsilon\,  v }{m_H^2} g_{\tt N-H} \then \sigma_0 = \inv{8\pi \mbe^2} \left[ \frac{\mbe m_N}{\mbe + m_N} \, \frac{\epsilon\, g_{\tt N-H} v}{\msm^2} \right]^2\,,
\eeq
where $v$ denotes the SM \vev, $m_N$ the nucleon mass, and $g_{\tt N-H} \simeq 0.0034 $ the Higgs-nucleon coupling \cite{Shifman:1978zn,Dawson:1989gr,Jungman:1995df}. 

For the range of parameters we consider, the temperature of the Bose gas at present, $T\be $ is very small, so that 
\bal
\sigma &= \frac{\epsilon^2}{8\pi \mbe^2} \left( \frac{\mbe/m_N}{1 + \mbe/m_N} \, \frac{g_{\tt N-H} v\, m_N}{\msm^2} \right)^2\, \left(1 +  r^2 \frac{T\be}{\mbe\mbox{\tt v}^2}  \right) \,, \mcr
& \simeq 6.93 \times 10^{-34}  \left(\frac\epsilon{1 + \mbe/m_N} \right)^2\, \left(1 +  \frac{m_N^3}{\mbe^3} \frac{T\be}{600\, ^oK}  \right)\, \cm^2\,,
\label{eq:cs}
\end{align}
where $r$ is defined in \cref{eq:r.def},  {\tt v}$\simeq10^{-3}$ is the nucleon-dark matter relative velocity and, as above, $ r = \msm/\mbe$.

\begin{figure}[ht]
$$ 
\includegraphics[height=2.5in]{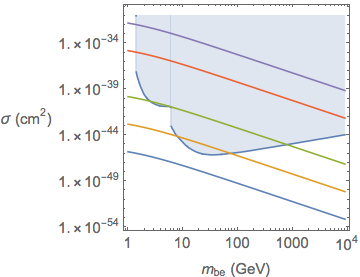} \qquad
\includegraphics[height=2.5in]{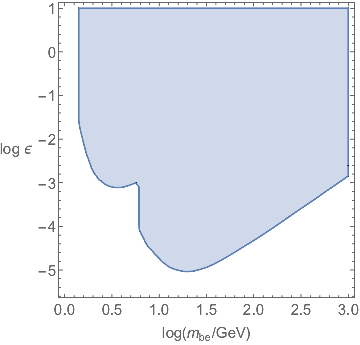} 
$$
\caption{Left: the curves give the direct-detection cross section \cref{eq:cs} for (lower to upper curves, respectively) $ \log\epsilon=-6,\,-4.5,\,-3,\,-0.5,\,1 $ with the shaded area denoting the region excluded by the XENON and CDMSLite  experiments. Right: the shaded area denotes the region of the $ \mbe - \epsilon $ plane excluded by direct-detection.}
\label{fig:cs}
\end{figure}

These results can be compared to the most recent XENON \cite{Aprile:2017iyp} and CDMSLite \cite{Agnese:2017jvy} constraints, we present the results in Fig.\ref{fig:cs}. We find that the leading temperature correction in \cref{eq:cs} is negligible except for very small $ \mbe $, in this case, however the cross section itself is very small.

The graphs in Fig. \ref{fig:cs} represent the strongest constraints on the model parameters. If the parameters are allowed by the direct-detection constraint the model will satisfy the relic abundance requirement for an appropriate choice of $ \mu $.

\section{Bose condensate in the small mass region}
\label{sec:BEC}
As noted above, a condensate can occur when the gas has sub-eV masses. In this case, however, there are additional constraints stemming form the possible effects of such light particles on large scale structure (LSS) formation and on big-bang nucleosynthesis (BBN). In this section we will investigate the  regions in parameter space  allowed by these constraints assuming that the gas is currently condensed; as noted in section \ref{sec:cosmology} this ensures the presence of a condensate in earlier times~\footnote{At least as long as $ \x > \lbe/8.8$, see \cref{eq:x.lim}.}. 

For the small masses needed to ensure the presence of a BEc now (see below) the condition $H = \Gamma $ used in section \ref{sec:width} (\cref{eq:Td,eq:Gamma}) would require a coupling $ \eps $ orders of magnitude above the perturbativity limit~\footnote{To see this we used \cref{eq:g44,eq:g42,eq:g24,eq:g22} since the expressions in \cref{eq:NRG} are not valid for the small values of $ \mbe $ considered here.} (see sect. \ref{sec:introduction}), hence in this case the gas is decoupled from the SM during the BBN and LSS epochs.

\begin{figure}[ht]
$$
\includegraphics[height=3in]{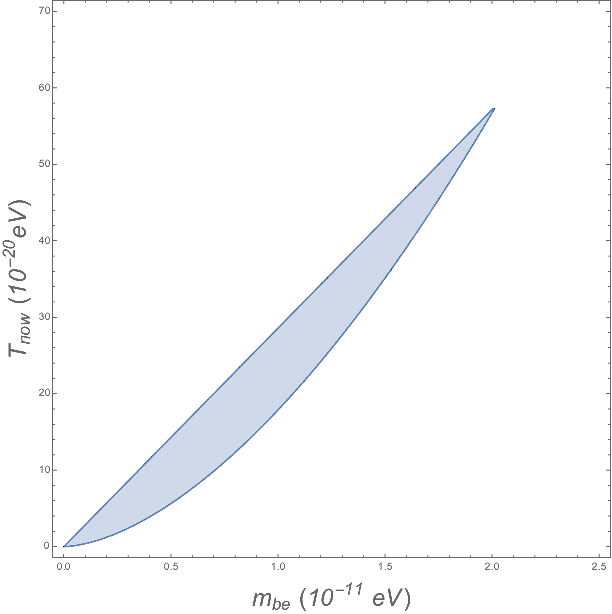}\qquad
\includegraphics[height=3in]{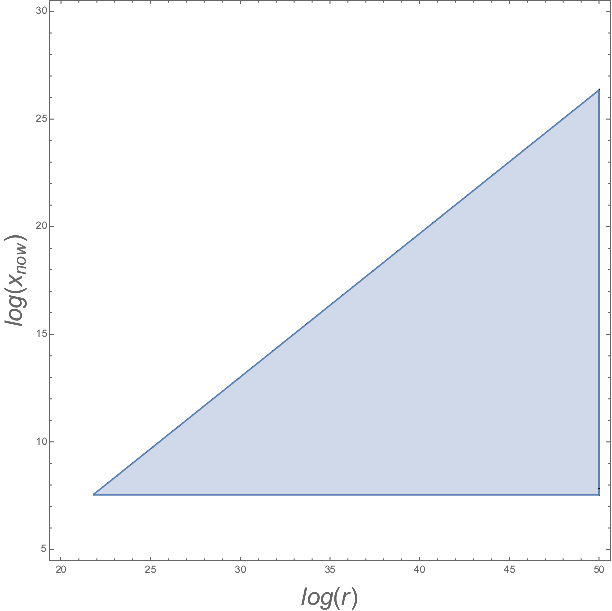}
$$
\caption{Regions of the $ \mbe-T $ and $ r-\x$ planes where a non-relativistic Bose condensate occurs consistent with the LSS constraint of \cref{eq:LSS}. On the left-hand graph the low-$T$ limit results form \cref{eq:BEC-reg}, while the upper limit is due to \cref{eq:LSS}.}
\label{fig:LSS}
\end{figure}

LSS formation  occurred at redshift $ z_{\tt LSS}\sim 3400$, when the matter-dominated era began \cite{Kolb:1990vq}. To ensure that the Bose gas does not interfere with the formation of structure we require it to be non-relativistic at that time; in addition, since we  assume the presence of a BEc at present, a BEc was also present at the LSS epoch (sect. \ref{sect:condensate}). Then the conservation of $ a^3 \sbe$ gives, using \cref{eq:nrel.rns}, $a^3 \x^{-3/2} =$ constant ($a$ denotes the scale factor in the Robertson-Walker metric); equivalently,
\beq 
\left. \frac{a^2}\x \right|_{\tt now} = \left. \frac{a^2}\x \right|_{\tt LSS} \then 
\x_{\tt now} = \left(1+z_{\tt LSS}\right)^2 \x_{\tt LSS}\,.
\eeq
Since the gas must be non-relativistic during the LSS epoch, $ \x_{\tt LSS}> 3 $, so we have
\beq
\x_{\tt now} > 3.5 \times  10^7\,.
\label{eq:LSS}
\eeq
In addition, the requirement that a BEc be present now implies 
\beq
\frac{0.4\, \ev}\mbe \left. \ssm \right|_{\tt now}>  \left( \frac{ \mbe^2 }{2\pi \x_{\tt now}} \right)^{3/2} \zeta_{3/2}\,,
\label{eq:BEC-reg}
\eeq
where we used the fact that the gas is currently non-relativistic~\footnote{The $O(\lbe)$ corrections can be ignored in this case, see appendix \ref{sec:app.A}.}.

The regions in the $ \mbe-T$ and $ \mbe -\x $ planes allowed by \cref{eq:LSS} and \cref{eq:BEC-reg} are given in Figure \ref{fig:LSS} (here $T$ refers to the gas temperature). It is worth noting that if these conditions occur at present, most of the gas will be in the condensate: using \cref{eq:rel.ab} and \cref{eq:LSS} the gas fraction in the excited states is given by
\beq
\left. \frac{\ch\up e}{\ch} \right|_{\tt now} < \left( \frac\mbe{1.82\, \ev} \right)^4\,,
\eeq
which is negligible in view of the range of masses being here considered (see figure \ref{fig:LSS}).

\bigskip \bigskip

We now turn to the BBN constraints. We write the contributions from the gas to the energy density in the form of an effective number of neutrino species $ \Delta N_\nu $:
\beq
\rho\be|_{\tt BBN} = \frac3{\pi^2} \frac74 \left( \frac 4{11} \right)^{4/3} \!\! \Delta N_\nu \, T_\gamma^4 \simeq 0.138 \Delta N_\nu\, T_\gamma^4\,,
\eeq
where $ T_\gamma \simeq 0.06 \, \mev $ denotes the photon temperature during BBN \cite{Baumann}. Imposing the relic-abundance constraint \cref{eq:rel.ab} we find, using \cref{eq:nu} and \cref{eq:rbe-sbe},
\beq
\Delta N_\nu = 7.2 \times 10^{-5} + 7.24 \frac{\mbe^4}{T_\gamma^4} \left[ r\be(\x_{\tt BBN}) - \nu\be(\x_{\tt BBN}) \right]_{\delta  \ge \fBB} \,.
\eeq
where $r\be - \nu\be$ corresponds to the energy outside the condensate. 

The limit (see \cite{Cyburt:2015mya}) $ -0.7< \Delta N_\nu < 0.4 $ shows that the first contribution to $ \Delta  N_\nu $ can be ignored. Also, the LSS constraint $ \mbe < 2 \times 10^{-11} \, \ev $ (see Fig. \ref{fig:LSS}), implies $ ( \mbe / T_\gamma) \lesssim 10^{-62} $, so that the second contribution to $ \Delta N_\nu $ is also small except if the gas was ultra-relativistic during BBN. In this case
\beq
\Delta N_\nu \simeq 4.76 \left( \frac{\mbe }{T_\gamma \x_{\tt BBN} } \right)^4 \left[ 1 +  \frac{5\lbe}{16 \pi^2} \right]\,, \quad \x_{\tt BBN} \ll 1\,,
\label{eq:Nnu}
\eeq
so the BBN constraint is significant only in the extreme ultra-relativistic case where $ x_{\tt BBN} <  10^{-62} $.

\begin{figure}[ht]
$$
\includegraphics[height=3in]{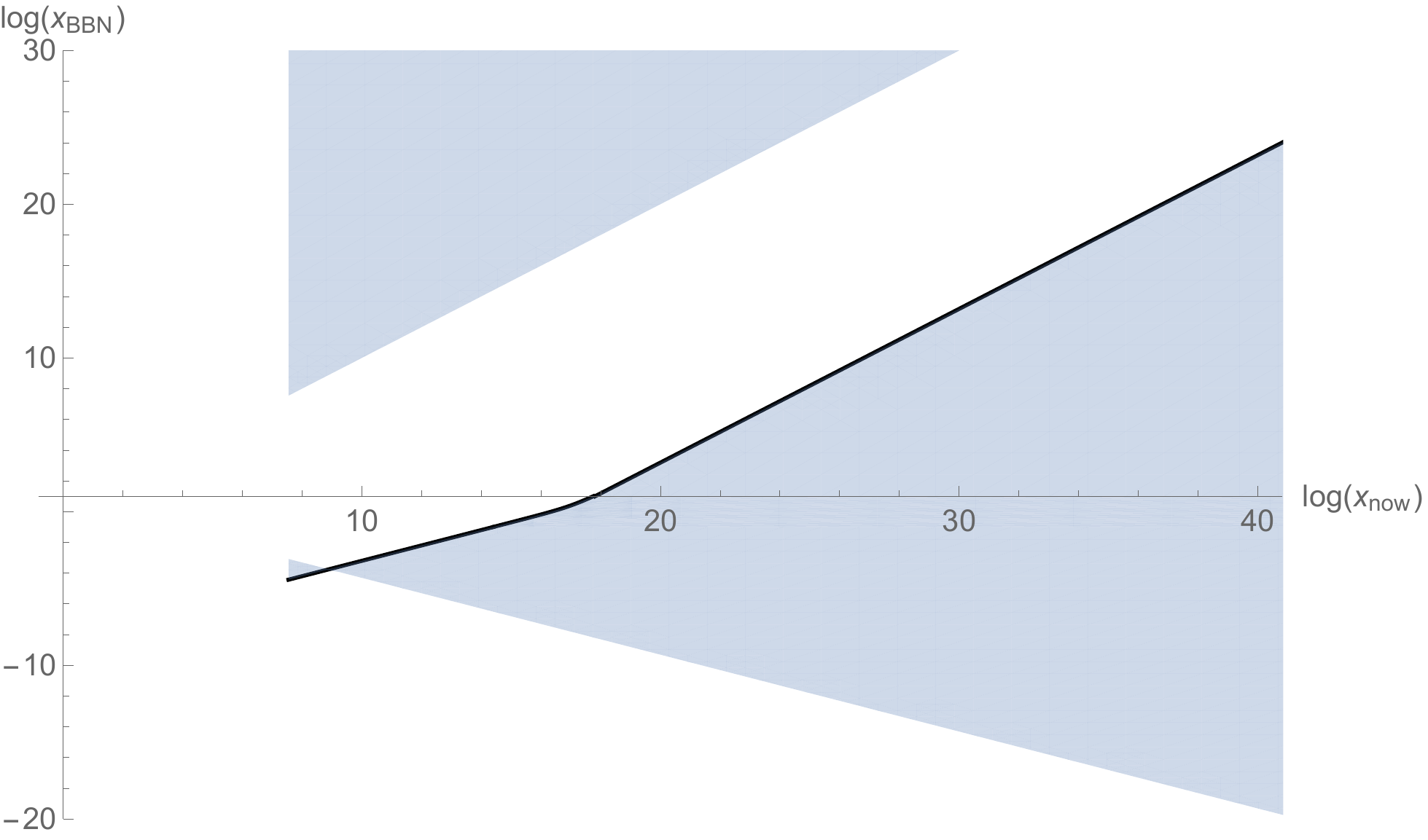}
$$
\caption{Region in the $ \x_{\tt BBN} - \x_{\tt now} $ plane consistent with the conservation laws, and with the assumption that a BEc is currently present. We used the expressions in appendix \ref{sec:app.A} and $ \ssm|_{\tt now} = 2889.2/$cm$^3$, $ \ssm|_{\tt BBN} = 4.82 \times 10^{28}/$cm$^3$ and took $ \lbe = 0.5$. When $ \lbe =0 $ the allowed region collapses to the bold dark line in the figure.
}
\label{fig:xBBN-xnow}
\end{figure}

To examine this possibility we first obtain in figure \ref{fig:xBBN-xnow} the regions in the $ \x_{\tt BBN} - \x_{\tt now} $ plane consistent with the fact that $ \sbe/\ssm$ and $ \ch/\ssm$ are conserved, together with the assumption  that a BEc is currently present. The lower bound in this region corresponds to $\x_{\tt BBN}  \ge  4.9/\sqrt{x_{\tt now}}$; using this, and the BBN constraint $ \Delta N_\nu < 0.4 $  in \cref{eq:Nnu}, we  obtain
\beq
x_{\tt now} < 1.1 \times   10^{125} \left( \frac \mbe{ 10^{-11} \ev} \right)^{-2}\left(1 - \frac{5 \lbe}{32\pi^2} \right) \,,
\label{eq:BBN}
\eeq
To understand the gap that appears in  figure \ref{fig:xBBN-xnow} consider the expressions in \cref{eq:qs_bec}: we write  $s\be = s\be\up e + [\lbe \bec^2/(2 \mbe^2)] s\be\up c $ (this defines~\footnote{By definition, $s\be\up c$ contains all terms $ \propto\bec^2 $ (up to a factor of $ \lbe/(2\mbe)$) in \cref{eq:qs_bec}; $s\be\up e $ contains all remaining terms. Note that $s\be\up e $ includes $O( \lbe)$ contirbutions.} $ s\be\up{e,c}$) and use $\bec^2  = [ \ch - \ch\up e]/\mbe +O(\lbe)  $; then, noting that $ s\be\up e \gg s\be\up c \ch \up e $ (which we verified numerically), and using the fact that $ s\be/s\sm$ and $ \ch/s\sm$ are constant, we find
\beq
\frac{\left[ s\be\up e/s\sm \right]_{\tt BBN} -  \left[s\be\up e/s\sm \right]_{\tt now}}{\left[ s\be\up c \right]_{\tt now} -  \left[s\be\up c\right]_{\tt BBN}} = \frac\lbe{2\mbe^3} \frac\ch{s\sm} >  \left.  \frac\lbe{2\mbe^3} \frac{\ch\up e}{s\sm} \right|_{\tt now}\,,
\eeq
where the inequality on the right-hand side  imposes the constraint that a BEc is present now. The  gap in figure \ref{fig:xBBN-xnow} corresponds to values of $x_{\tt BBN,\, now} $ where the denominator and numerator have opposite signs. For example, if the gas is non-relativistic during nucleosynthesis, 
\beq
\frac{1 - \vartheta \left( x_{\tt now}/x_{\tt BBN} \right)^{3/2}}{1 - \sqrt{x_{\tt now}/x_{\tt BBN} }} > \frac{3 \lbe}{40\pi} \frac{\zeta_{3/2}^2}{\zeta_{5/2}} \inv{2\pi x_{\tt now}}\,, \quad \vartheta = \frac{s\sm|_{\tt now}}{s\sm|_{\tt BBN}} \simeq 6 \times 10^{-26}\,;
\eeq
in this case the gap corresponds to $ \log x_{\tt now} \gesim \log x_{\tt BBN} \gesim -16.8 +  \log x_{\tt now} $.

The parameter region where the gas exhibits a BEc now and satisfies both the LSS and BBN constraints are determined by \cref{eq:BBN}, \cref{eq:LSS} and the allowed $ \x_{\tt BBN} - \x_{\tt now} $ and $ \mbe - T_{\tt now} $ regions in figures \ref{fig:LSS} and \ref{fig:xBBN-xnow}, respectively. It is worth noting that when $ \lbe=0 $ the allowed region in the $ \x_{\tt BBN} - \x_{\tt now} $ plane reduces to the dark line in figure \ref{fig:xBBN-xnow}, in which case the BBN constraint does not impose new restrictions.

\bigskip \bigskip

It remains to see whether a gas satisfying \cref{eq:LSS} can be in equilibrium with the SM at an epoch earlier than that of BBN. Given the small range for $ \mbe $ and the large values of $ \x_{\tt now} $, such equilibrium could have occurred only when the gas was  ultra-relativistic, in which environment the presence or absence of a condensate will have no effect. The situation then reduces to that of a standard Higgs-portal model with DM masses in the pico-eV range. Concerning direct detection experiments it is clear that for the very small masses being considered in this section the cross sections will be negligible. We will not consider these points further. 

\section{Comments and conclusions}
\label{sec:comments}

In this paper we investigated various properties of  a complex scalar model of dark matter and studied the possible presence of a Bose condensate, which can occur even in the relativistic regime due to the presence of a conserved charge, associated an exact  ``dark'' $ \ui $ symmetry.

We showed that a Bose condensate will be present at sufficiently early times provided the charge per unit entropy is above a $ \lbe$ and $ \mbe $-dependent minimum (when $ \mbe > \msm $ this minimum will also depend on $ \epsilon $); for $ \lbe =0 $ a condensate will always form in the early universe. As $ T \to \infty $ one-loop results  suggest that the condensate will disappear despite the vanishing of the co-moving volume in that limit. The constraints derived form large scale structure formation imply that a condensate will persist until the present only  if the dark matter mass is in the pico-eV range. 

The model can meet the relic-density constraint for all masses in the cold dark-matter regime  ($ \mbe \gtrsim 1\, \gev$) provided the portal coupling $ \epsilon \le 0.1 $ and for a wide range of masses; for larger values of $ \epsilon $ the mass range is somewhat narrower, see Fig. \ref{fig:m-T}). The limits derived from direct-detection experiments are much more restrictive allowing only small couplings and/or small masses (Fig. \ref{fig:cs}); still the allowed region in parameter space is considerably extended compared to the usual Higgs-portal model \cite{Athron:2017kgt} because of the presence of a chemical potential that can be adjusted to ensure the correct relic density.

For WIMP-like masses we have shown above that there is no condensate for $ T < T_d $ but that a condensate can form in the early universe, at least for a period of time; at very high temperatures the condensate then carries the net charge of the gas, but most of the energy density is carried by the excited states (section \ref{sec:cosmology}). In contrast, for very small masses, $ \mbe \sim 10^{-12}\, \ev $  the gas can form a condensate even at present temperatures, while also satisfying the relic abundance requirement. In this case, however, the Bose gas and the SM are never in equilibrium  (assuming natural values  of the portal coupling $ \epsilon $).

Most of the radiative effects in this model are small, being suppressed not only by powers of $ \lbe$, but, in the non-relativistic limit, by inverse powers of $ \mbe/T$. We found two exceptions: first, the above-mentioned condition on the formation of a condensate in the early universe. Second, the constraint in \cref{eq:BBN} derived from BBN.

We have not discussed indirect detection constraints because, for WIMP-like masses they will be identical to those derived for the standard Higgs portal models \cite{TheFermi-LAT:2017vmf}.

\appendix

\section{Thermodynamics of a Bose gas}
\label{sec:app.A}
In this appendix we provide for completeness a summary of the Bose gas thermodynamics. We begin with the Lagrangian
\beq
\lcal = |\partial \chi|^2 - m^2 |\chi|^2 - \half \lbe |\chi|^4\,,
\eeq
and write $ \chi = (A_1+ i A_2)/\sqrt{2} $. Then the Hamiltonian and total conserved charge $\Ch$ are given by
\beq
H = \int d^3\xx \left[ \half \pibf^2  +\half |\nabla \AA|^2  + V \right]\,, \qquad
\Ch = -\int d^3\xx \left( A_1 \pi_2 - A_2 \pi_1 \right)\,,
\eeq
where $ \pi_i $ is the canonical momentum conjugate to $A_i$.

To include the possibility of a Bose condensate we replace $ A_1 \to A_1 + \bec $; using then standard techniques of finite-temperature field theory (we use here the Matsubara formalism) \cite{Kapusta} we find that to $O(\lbe) $ the  pressure $P\be$ is given by \cite{Kapusta:1981aa,Haber:1981ts}
\beq
P\be = \frac{\mu^2 - \mbe^2}2 \bec^2 + \frac23 \int \dtp \, p^2 \fcal_+ + \inv8\lbe \bec^4  - \lbe \left( \half \bec^2 + \int \dtp \, \fcal_+ \right)^2 +O(\lbe^2) \,,
 \eeq
 where
\begin{align}
\fcal_\pm &= \inv{e^{\beta (E- \mu)} -1}  \pm \inv{ e^{\beta (E+\mu)} -1}\,; \quad  \bar\fcal_\pm = \left. \fcal_\pm \right|_{\mu=\mbe}\,, \mcr
\dtp &= \frac{d^3\pp}{(2\pi)^3\, 2 E}\,; \quad E = \sqrt{\pp^2+\mbe^2}\,.
\label{eq:some_defs}
\end{align}

When one adds the coupling $ \epsilon |\phi|^2 |\chi|^2$ to the Standard Model (see \cref{eq:model}) there is an additional contribution
\beq
\Delta P\be = - \epsilon F_{\tt H}\left( \half \bec^2 + \int \dtp \, \fcal_+ \right) \,; \quad F_{\tt H}=\frac{\msm^2}{\pi^2}\int_0^\infty d\alpha \frac{\sinh^2\alpha }{e^{(\msm/T) \cosh\alpha}-1} \,,
\label{eq:Psmdm}
\eeq
where $F_{\tt H}$ is generated by the $ \phi $, when the $ \phi $ acquires an expectation value  $ F_{\tt H} \to v^2 + F_{\tt H}/4$. This term is subdominant when $ \msm > \mbe $ as we will assume for the most part of this paper; note also that stability conditions (see section \ref{sec:introduction}) do not allow $ \epsilon  $ to be too large and negative. The {\em total} pressure has additional terms, generated by the standard model; these terms, however, do not involve $\bec $.

Before proceeding we remark on the type of perturbative expansion we will use: we assume  that $\bec$ is independent of $ \lbe $, and $ \mu $ to have a $\lbe $ dependence~\footnote{If, on the other hand $ \mu $ is assumed to be independent of $ \lbe $, then $ \bec \propto 1/\sqrt{\lbe} $ diverges as $ \lbe \to 0 $.}; we believe this to be reasonable because, for example, the condition for the presence of a BEc when $ \lbe =0$ is $ \mu = \mbe$, and becomes $ \mu > \mbe $ when $ \lbe \not=0$ (see below) that naturally leads to a relation of the form  $ \mu = \mbe + O(\lbe) $.

The zero-momentum component $\bec$ is determined by the condition that it minimizes the thermodynamic potential $-P\be(\bec,\mu,T) $:
\beq 
\deriva{P\be}\bec{} = \lbe \bec \left\{ \delta - \fBB - \half \bec^2  \right\} + O(\lbe^2)\,,
\label{eq:curv}
\eeq
where ($ \bar\fcal_\pm $ are defined in \cref{{eq:some_defs}})
\beq
\mu^2 = \mbe^2 + \lbe \delta\,; \qquad \fBB = 2\int \dtp \, \bar\fcal_+\,.
\label{eq:del.F}
\eeq
So there are two cases:
\ben
\item $ \delta < \fBB $: then there's a single extremum, $ \bec =0 $, which is a maximum and corresponds to the stable state; there is no BEc. 

\item $ \delta > \fBB $: then there are two extrema: $ \bec =0 $ which is now a minimum, and does not correspond to the stable state, and 
\beq
 \bec^2 = 2 \left( \delta - \fBB \right)  + O(\lbe)\,,
 \eeq
 which is a maximum and corresponds to the stable (BEc) configuration.
\een

The transition occurs when $ \delta = \fBB $; approximating $ \fBB \simeq \fBB(  \mbe  =0) $ we find that the critical temperature is
\beq
T_c^2 \simeq \frac6\lbe(\mu^2 - \mbe^2)\,,
\eeq
which is a known result \cite{Kapusta:1981aa,Haber:1981ts}.

From $P\be$ we find the expressions for the charge density $\ch$ and entropy density $\sbe$  to $O(\lbe )$:
\bit
\item $ \delta < \fBB$:
\bal
P\be &=  \frac23 \int \dtp \, p^2\fcal_+   -  \lbe \left( \int \dtp \, \fcal_+ \right)^2 \mcr
\ch &=  \int \dpp \,  \fcal_-  - \lbe \left( \int \dtp \, \fcal_+ \right) \left( \int \dpp \, p^{-2} \fcal_- \right)\mcr
\sbe &=  \int \dpp \left( 1 - \lbe \frac{K^2}{p^2}\right) \sum_\pm\left[ (n\be^\pm+1) \ln( n\be^\pm+1) - n\be^\pm \ln n\be^\pm \right] \,,
\label{eq:qs_nobec}
\end{align}
where $K^2 =   4\int \dtp \, \fcal_+$.
\item $ \delta = \fBB $:
\bal
P\be &=\frac23 \int d\tilde p \, p^2 \bar\fcal_+  - \inv4\lbe \fBB \left( \fBB- \frac2m \int d_3\pp \bar\fcal_-  \right) \,, \mcr
\ch &= \int \dpp \bar\fcal_- + \frac{4\lbe\fBB}m   \left(\frac m4 \int \dpp \frac{\bar\fcal_+ - \bar\fcal_-}{p^2} +\int \dtp \, \frac{E + m/2}{E + m} \bar\fcal_+ \right) \,,\mcr
\sbe &= \int d_3\pp \left( 1 - \lbe \frac{2\fBB}{p^2}\right) \sum_\pm\left[ (n\be^\pm+1) \ln( n\be^\pm+1) - n\be^\pm \ln n\be^\pm \right]_{\mu = \mbe} \cr
& \phantom{s=}\quad + \frac{\lbe \fBB}T \int d\tilde p \left\{ \frac{E^2+p^2}{p^2} \left( \bar\fcal_- - \bar\fcal_+ \right) + \frac{3E^2+ m E - m^2}{m(E+m)} \bar\fcal_- \right\} 
\,.
\label{eq:qs_bec}
\end{align}

\item $ \delta > \fBB $:
\bal
P\be &=  \frac23 \int d\tilde p \, p^2 \bar\fcal_+ - \inv 4 \lbe \left[\fBB^2 - \frac{\bec^4}2 -  \frac{\bec^2+2\fBB}m \int d_3\pp \bar\fcal_-  \right]\,,
\mcr
\ch &= \ch\up c + \int \dpp \bar\fcal_- + O(\lbe)\,,\mcr
\sbe &= \int d_3\pp \left( 1 - \lbe \frac{2(\bec^2 + \fBB)}{p^2}\right) \sum_\pm\left[ (n\be^\pm+1) \ln( n\be^\pm+1) - n\be^\pm \ln n\be^\pm \right]_{\mu = \mbe} \cr
& \phantom{s=}\quad + \frac{\lbe (\fBB + \bec^2/2)}T \int d\tilde p \left\{ \frac{E^2+p^2}{p^2} \left( \bar\fcal_- - \bar\fcal_+ \right) + \frac{3E^2+ m E - m^2}{m(E+m)} \bar\fcal_- \right\}\,.
\label{eq:qs_bec}
\end{align}
with $ \ch\up c = \mbe \bec^2 + O(\lbe)$. The $O(\lbe)$ corrections to $ \ch $ in the BEc phase are obtained from the $ O(\lbe^2) $ terms in $ P\be $, fortunately these are not needed.
\eit

\begin{figure}[t]
$$
\includegraphics[width=3in]{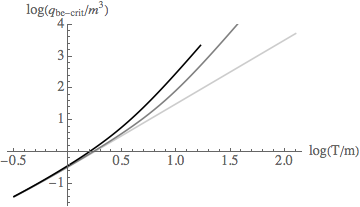} 
$$
\caption{Plot of the critical density as a function of $T$ for $ \lbe = 0 $ (light gray), $ 0.1$ (dark gray) and $0.5$ (black).}
\label{fig:crit}
\end{figure}

The curvature  of the thermodynamic potential $-P\be(\bec,\mu,T) $ at $\bec=0$ equals $ \lbe( \fBB - \delta)  \simeq  \lbe T^2/6 + \mbe^2 - \mu^2$ for large $T$ (see \cref{eq:curv}). In this regime the radiative corrections oppose the formation of a condensate; if this is indicative of the exact result, the condensate will disappear  as $ T \to \infty $. The behavior of the critical density ($\ch$ at the transition) is given in Fig. \ref{fig:crit} which also illustrates the effects of the $O(\lbe T^2)$ contributions.

\begin{figure}[t]
$$
\includegraphics[width=3in]{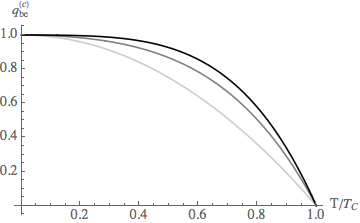} 
$$
\caption{Plot of the condensate density $ \ch\up c$ as a function of $T$ for constant volume and $ \lbe = 0 $ (light gray), $ 0.1$ (dark gray) and $0.5$ (black), when the critical temperature (see text) $T_C = 10 \mbe $. When $ T_C \ll \mbe $ the $O(\lbe ) $ effects are negligible.}
\label{fig:bec-T}
\end{figure}

When the volume $V$ is constant and the total charge in the system is $ \Ch$ the behavior of the condensate as a function of $T$ can be obtained using standard arguments; the results are illustrated in Fig. \ref{fig:bec-T} where the critical temperature $T_C$ is defined by requiring $ \ch = \Ch/V $ when $ \delta = \fBB $.

In the non-relativistic limit ($ \x \gg1 $) the $ O(\lbe )$ can be ignored in the phase where there is no condensate. To see this, consider, for example the expression for $P\be$:
\beq
P\be = \frac{\mbe^4}{\pi^2 \x^2 } \left[ \cosh(\beta\mu)  K_2(\x) + \frac{\cosh(2\beta\mu)}4 K_2(2\x) - \frac{\lbe \cosh^2(\beta\mu)}{4\pi^2}  K_1^2(\x) + \cdots \right]\,, 
\label{eq:Pnr}
\eeq
which shows that the leading $ O(\lbe)$ corrections are smaller than the subdominant $ O(\lbe^0)$ contributions. This behavior is reproduced in all thermodynamic quantities in when $ \x\gg1$ and there is no BEc.

We also need the behavior of the thermodynamic quantities at the transition (when $ \delta = \fBB $) in the ultra-relativistic ($ \x\ll1$) and non-relativistic ($\x\gg1$) limits:
\bal
\x\ll1:\quad P\be &= \frac{\pi^2 \mbe^4}{45\, \x^4} \left[ 1 +  \frac{15\lbe}{16 \pi^2} \right] + \cdots \mcr
\ch & = \frac{\mbe^3}{3 \,\x^2} \left[ 1 - \frac{3\x}{\pi^2} + \frac\lbe{12\, \x^2} \left( 1 - \frac3{\pi^2} \x \ln \x \right) +  \cdots \right] \mcr
\sbe &= \frac{4\pi^2 \mbe^3}{45\, \x^3} \left[ 1 +  \frac{5\lbe}{16 \pi^2} \right] + \cdots \mcr
\rho\be &=\frac{\pi^2 \mbe^4}{15\, \x^4} \left[ 1 +  \frac{5\lbe}{16 \pi^2} \right] + \cdots 
 \label{eq:rel.rns} \\[6pt]
\x\gg1: \quad P\be &=  \frac{\mbe^4\zeta_{5/2}}{(2\pi)^{3/2}  \x^{5/2}}   \left[ 1 +   \frac{\zeta_{7/2}}{\zeta_{5/2}} \frac{15 }{8\x}  + \cdots \right]  + \lbe   \frac{\mbe^4\zeta^2_{3/2}}{(2 \pi \x)^3}   + \cdots   \mcr
\ch &=  \frac{\mbe^3 \zeta_{3/2}}{(2\pi \x)^{3/2}} \left[ 1 +  \frac{\zeta_{5/2}}{\zeta_{3/2}}\frac{15}{8 \x} +  \cdots \right] + \frac{3\lbe \mbe^3\zeta_{3/2}^2}{2(2\pi \x)^3} + \cdots  \mcr
\sbe &=\frac{5 \mbe^3 \zeta_{5/2}}{2(2\pi \x)^{3/2}} \left[ 1 + \frac{\zeta_{7/2}}{\zeta_{5/2}} \frac{21}{8\x} + \cdots \right] + \frac{9\lbe \mbe^3 \zeta_{3/2}\zeta_{5/2}}{128\pi^3 \x^3} + \cdots \mcr
\rho\be &= \frac{\mbe^4 \zeta_{3/2}}{(2\pi \x)^{3/2} } \left[ 1 + \frac{\zeta_{5/2}}{\zeta_{3/2}} \frac{27}{8\x} + \cdots \right] + \frac{ \lbe \mbe^4 \zeta_{3/2}^2}{(2\pi \x)^3} + \cdots
\label{eq:nrel.rns}
\end{align}
where $ \rho\be$ is the energy density.

In particular, for small $\x$,
\beq
\frac{\ch}{\sbe} = \frac{15}{4\pi^2  } \left[ 1 -  \frac{5\lbe}{8\pi^2} \right]  \left[ \x - \frac{3\x^2}{\pi^2} + \frac\lbe{12\, \x} \left( 1 - \frac3{\pi^2} \x \ln \x \right) +  \cdots \right] \,, \quad (\delta=\fBB,\, \x < 1 )
\eeq
which has a minimum when
\beq
\x_{\tt min} = \sqrt{\frac\lbe{12}} + \frac{3\lbe}{8\pi^2} + \cdots
\eeq

The above minimum occurs when the $O(\lbe)$ corrections to $ \ch $ are of the same size as the $O(\lbe^0)$ contributions, so the validity of the expressions for such values of $\x$ should be examined.  The leading expression for $ \ch $ is $\propto \int d^3\pp \bar\fcal_- $ and behaves as $ \x^{-2} $, instead of $ \x^{-3}$ as might be expected on dimensional grounds; such a suppression is not present in the $ O(\lbe)$ corrections. We argue that a reasonable estimate of the region where perturbation theory is valid is obtained by comparing the $ O(\lbe)$ corrections to $ \ch$ with a quantity that does not exhibit the above suppression, such as $ \int d^3\pp \bar\fcal_+ $. Using this we obtain
\beq
\int \dpp \bar\fcal_+ > \frac{\mbe^3\lbe}{36 \,\x^4}   \left( 1 - \frac3{\pi^2} \x \ln \x   +  \cdots \right) \then \frac{\x}{1 - (3/\pi^2) \x\ln \x} > \frac{\lbe}{8.8}
\label{eq:x.lim}
\eeq
as specifying the lowest value of $ \x$ for which our perturbative expressions are trustworthy. Since $ \x_{\tt min} $ satisfies this condition, the expression for $ \ch/\sbe $ can be trusted near the minimum.

\subsection{$\chi$ propagator.}

The above Hamiltonian and charge operators can be used to derive the propagator and Feynman rules in the fninte-temperature real-time formalism, which we use in some of our calculations. Defining, as usual~\footnote{We follow the conventions of LeBellac \cite{Bellac:2011kqa}}
\beq
D^>_{ij}(x - x') =  \vevof{ A_i(x) A_j(x')}_\beta\,,\quad D^<_{ij}(x - x') = \vevof{ A_j(x')A_i(x) }_\beta\,,
\eeq
(so that $D^<_{ij}(x - x') = D^>_{ji}(x' - x)$)
where
\beq
\vevof\cdots_\beta = \frac{ \tr{e^{ - \beta H} \cdots} }{\tr{e^{ - \beta H}}}\,.
\eeq
Then if,
\beq
\rho_{ij}(k) = D^>_{ij}(k) - D^<_{ij}(k)\,; \quad D^\gtrless_{ij} (k) = \int d^4 x\, e^{+ i k.x} D^\gtrless_{ij}(x)\,,
\eeq
we have
\beq
D^<_{ij}(k)= f(k_0) \rho_{ij}(k) \,, \quad D^>_{ij}(k) = - f(-k_0)  \rho_{ij}(k)\,; \qquad f(k_0) = \left( e^{k_0 \beta } - 1 \right)^{-1}\,.
\label{eq:D.f}
\eeq
A straightforward (though tedious) calculation yields
\begin{align}
\rho(k) =& 2 \pi \varepsilon(k_0) \left[ \frac{ \delta( \omega^2 - \Omega_+^2) - \delta(\omega^2 - \Omega_-^2)}{\Omega_+^2 - \Omega_-^2} \right] \, \rBB(k)\,,\mcr
\rBB(k)=&{\bpm k^2 + \mu^2 - m^2 - \lbe \bec^2/2 & - 2 i \mu k_0 \cr  2 i \mu k_0 & k^2 + \mu^2 - m^2 - 3\lbe \bec^2/2 \epm} \,.
\end{align}
This has the expected form when $ \mu =0 $. For the calculations in this paper we only need the expression when $ \lbe =0 $:
\beq
\rho(k)|_{\lbe=0} = \pi \sum_{s=\pm1} ( \mati \pm \tau_2)   \varepsilon(k_0 \mp \mu) \delta( (k_0 \mp \mu)^2 - \eb_\kk^2) \,,
\label{eq:rhobe}
\eeq
where $ \eb_\kk = \sqrt{\mbe^2 + \kk^2}$. This expression is also valid in the presence of a condensate, when $ \mu = \mbe $.

\subsection{Higgs propagator and resonant contributions}
When the SM and the Bose gas are in thermal equilibrium a similar expression can be derived for the Higgs propagator, however, this approach misses an important resonant contribution which can occur when $ \msm = 2 \mbe $; to include it we replace
\beq
2 \pi \delta(p^2 - \msm^2) \to \frac{2\Gamma_{\tt H}\, \msm}{(p^2- \msm^2)^2 + ( \Gamma _{\tt H}\, \msm)^2}
\eeq
in $D^\gtrless_{\tt H}$, where $\Gamma_{\tt H}$ denotes the Higgs width.

\def\mred{m_{\tt red}}

\section{Appendix: Cross section in the presence of a condensate}
\label{sec:app.B}

In this case, writing again $ \chi \to [(A_1+\bec) + i A_2]/\sqrt{2} $ we find, to lowest order,
\bea
\vevof {W_{i \to f}}_\beta 
&=& \bec^2 \int d^4x \, d^4 y e^{-i (p - q).(x - y) }  \vevof{ T_c\left[A_1(t - i\beta ,\xx) A_1(y) \right]}_\beta \cr
&& \quad  + \inv4\int d^4x \, d^4 y e^{-i (p - q).(x - y) } \left[  \vevof{ T_c\left[\AA^2(t - i\beta ,\xx) \AA^2(y) \right]}_\beta 
-\vevof{\AA^2}_\beta^2 \right] \,,
\eea
where $\vevof {W_{i \to f}}$ is defined in \cref{eq:Wfi}, $ \vcal $ denotes the volume of space time, and we assumed that the incoming momentum $p$ of the SM particle is different form its outoging momentum $q$. Now, using \cref{eq:D.f} and \cref{eq:rhobe} we find
\bea
\inv\vcal \vevof {W_{i \to f}} =  \bec^2 D^>_{11}(P)|_{\mu = \mbe} + \frac{g^2 T f(-P_0)}{2\pi|\PP|} \ln \left| \frac{1 + \nbe (E_-)}{1+\nbe(E_+)} \frac{1+\nbeb(E_-)}{1+\nbeb(E_+)} \right|_{\mu = \mbe} \,,
\eea
where $ n\be\up\pm$ are defined in \cref{eq:shortcuts}, $ E_\pm $ in  \cref{eq:Wif.no.BEC}, and $ P = p-q $. Then
\bal
\sigma &= \sigma\up1  + \sigma\up2 \,,\mcr
\sigma\up1 &= \frac{\ch\up c}{2 \mbe |\pp| \ch} \int' \frac{d^3\qq}{2 E_\qq (2\pi)^3} D_{11}^>(P)|_{\mu = \mbe}\,; \quad E_\qq = \sqrt{\qq^2 + \msp^2} \,,\mcr
\sigma\up2 &= \inv{2 \ch |\pp| }\int' \frac{d^3\qq}{2 E_\qq (2\pi)^3}  \frac{g^2 T f(-P_0)}{2\pi|\PP|} \ln \left| \frac{1 + \nbe (E_-)}{1+\nbe(E_+)} \frac{1+\nbeb(E_-)}{1+\nbeb(E_+)} \right|_{\mu = \mbe} \,,
\end{align}
where $ E_\qq $ is the energy of the outgoing $\sp $, $\ch$ the number density of Bose gas particles, and  we used  $ \ch\up c = \mbe \bec^2 $ for the number density in the condensate; the prime indicates that the region $ p = q $ should be excluded. 

In the non-relativistic limit, and for $\mbe \not= \msp$, we find
\beq
 \sigma\up1 
= - \frac{T n_0/n}{ 32 \pi \mbe \pp^2} \ln \left|f(-\ecal_-) f(\ecal_+)\right|\,; \quad \ecal_\pm = \frac{ 2 \mbe \pp^2}{\mbe^2 + \msp^2 \pm 2 \mbe \bar E_\pp} \,\,,
\eeq
where $ \eb_\pp $ is defined in \cref{eq:E.n}, and $f$ in \cref{eq:D.f}. For $ T \to 0 $ (so that $ \ch\up c  \to \ch $ ) this reduces to the standard result, $ \sigma\up1 \to [16\pi(\mbe+\msp)^2]^{-1} $; also, $ \sigma \up1 > 0 $ for all parameters of interest.

The evaluation of $ \sigma \up 2 $ is more involved. We begin with the non-relativistic expression for $ E_\pm $:
\beq
E_\pm = \mbe + \inv{8 \mbe |\PP|^2} \left[  |\PP|^2 \mp \frac\mbe\msp (\pp^2 - \qq^2) \right]^2\,.
\eeq
Then, defining new integration variables
 \beq
 w = \frac{|\PP|}{|\pp|} \,, \quad z = \inv w\left( \frac{|\qq|^2}{|\pp|^2}  -1 \right) \frac\msp\mbe\,,
 \eeq
 we find
 \beq
\sigma\up2  = \frac{T |\pp|}{256\pi^3 \ch\, \mbe} \int_0^\infty dw\, w \int_{(w-2)\mbe/\msp}^{(w+2)\mbe/\msp} \frac{dz}{\exp\{ 4 \ell w z   \} - 1} \ln \left| \frac{1 - \exp\{- \ell( w + z )^2 \}} {1 - \exp\{- \ell( w - z )^2 \}} \right| \,,
\eeq
where $ \ell = \beta |\pp|^2/(8\mbe)$. This must be evaluated numerically for moderate values of $ \ell $, while for $ \ell \to \infty $, it gives \cref{eq:cs}.

\end{document}